\documentstyle[12pt,aasms4]{article}

\slugcomment{To appear in the Astronomical Journal}

\lefthead{Thompson et al.}
\righthead{NICMOS Observations of the Hubble Deep Field}

\begin{document}

\title{NICMOS OBSERVATIONS OF THE HUBBLE DEEP FIELD: OBSERVATIONS,\\
    DATA REDUCTION, AND GALAXY PHOTOMETRY}

\author{Rodger I. Thompson}
\affil{Steward Observatory, University of Arizona,
    Tucson, AZ 85721}

\author{Lisa J. Storrie-Lombardi and Ray J. Weymann}
\affil{Carnegie Observatories, Pasadena, CA 91101}

\author{Marcia J. Rieke, Glenn Schneider, Elizabeth Stobie and Dyer Lytle}
\affil{Steward Observatory, University of Arizona,
    Tucson, AZ 85721}

\begin{abstract}
This paper presents data obtained during the NICMOS Guaranteed Time Observations of a portion of the Hubble Deep Field.  The data are in a catalog format similar to the publication of the original WFPC2 Hubble Deep Field Program (\cite{will96}).  The catalog contains 342 objects in a $49.1 \times 48.4\arcsec$ subfield of the total observed field, 235 of which are considered coincident with objects in the WFPC2 catalog. The $3\sigma$  signal to noise level is at an aperture AB magnitude of approximately 28.8 at 1.6 microns.  The catalog sources, listed in order of right ascension, are selected to satisfy a limiting signal to noise criterion of greater than or equal to 2.5.  This introduces a few false detections into the catalog and users should take careful note of the completeness and reliability levels for the catalog discussed in Sections \ref{comp} and \ref{rel}.  The catalog also contains a test parameter indicating the results of half catalog tests and the degree of coincidence with the original WFPC2 catalog.  
\end{abstract}


\section{Introduction}
Deep observations with NICMOS, devoted to understanding the nature of galaxy formation and evolution along with information on cosmological parameters, have always been an important aspect of the NICMOS Instrument Development Team program since its inception in 1984.  After the WFPC2 Hubble Deep Field (HDF) program in late 1995 it became obvious that observations at longer wavelengths would greatly enhance the value of the existing data in addition to satisfying the original intent of deep observations.  The smaller field of view of the NICMOS instrument made it necessary to choose between deep observations of a portion of the HDF and a survey of the entire HDF at a brighter limiting magnitude.  With the advent of the HDF there existed a large disparity between the depth of the HDF and the depth of observations at near infrared wavelengths (\cite{con97}).  Since the majority of objects that might be at redshifts unobservable with the WFPC2 were expected to be relatively faint, the IDT decided to conduct a limited spatial survey to the faintest possible magnitude.  The results of a NICMOS General Observer HDF survey program (\cite{dic97}) will provide coverage over the entire HDF.

The NICMOS HDF program consists of 127 orbits out of a total of 553 orbits for the entire GTO program.  Table \ref{hdf} shows the distribution of orbits between the two filters and two grisms. An additional two orbits were dedicated to confirmation of guide star acquisition. Although the bulk of the orbits are dedicated to imaging, the large comoving volume for line observations available to the grisms is very appealing.  The small number of grism orbits shown in Table \ref{hdf} were intended as test cases to see if more GTO orbits should be transferred to this program.  In order to expedite the delivery of image data to the community at large we have not concentrated on the reduction of the grism data and it is not presented in this publication.  The grism data will be published in subsequent papers.  As discussed below the grism observations, however, significantly influenced the choice of the field for NICMOS imaging observations.

As with the \cite{will96} paper the purpose of this publication is a presentation of the data and analysis techniques rather than a discussion of the scientific content of the data.  Future papers will discuss several implications of the new data.  In the following we present the rationale for the observation methods, the methods for image production and source extraction, the catalog, and a discussion of the quality of the data in terms of signal to noise, completeness and reliability.  Note that all magnitudes quoted in this paper are in the AB system.

\section{FIELD SELECTION}
The decision to devote part of the observational time to grism observations limited field choices to regions of the HDF that are not dominated by large bright foreground galaxies.  The slitless dispersed spectra of these galaxies would overlap large areas of the field of view and reduce the number of spectral  observations of fainter galaxies.  Although some information on high redshift objects was available at the time of field selection, no effort was made to bias the field position to include the largest number of high redshift sources.  

The Space Telescope Science Institute decision to schedule the NICMOS HDF observations during the camera 3 campaign in January of 1998 determined the acceptable range of roll orientations.  This time period was not part of the Continuous Viewing Zone (CVZ) opportunity period, however, it did offer a larger fraction of truly dark observing time.  Given these constraints a field located roughly at the center of the WFPC2 chip 4 field offered the best observational opportunities.  The J2000.0 center position is $12^{\rm h}$ $36^{\rm m}$  $45.129^{\rm s}$, $+62\arcdeg$ $12\arcmin$ $15.55\arcsec$.  There is a relatively bright star ($AB_H$ of approximately 22.1  magnitude) near the center that provides an excellent fiducial location for the grism observations. The final orient of $261.851 \arcdeg$ is the result of fine tuning to obtain the best possible guide star orientation.

\section{FILTER AND CAMERA SELECTION}

\subsection{Camera Selection}
All of the NICMOS HDF data in this paper are from camera 3.  The wide field format of camera 3 made it the obvious choice for HDF observations.  The campaign also utilized cameras 1 and 2 with the same integration parameters as camera 3 but they were not in focus during operation.  Parallel observation of these cameras with camera 3 prevented the occurence of the faint artifacts, termed bars, which occur when the autoflush and imaging output timing patterns overlap.  Similar integration parameters for all cameras prevents the parallel cameras from defaulting into the autoflush pattern.  The data from these cameras may be useful for background characterization but are not analyzed in this publication.

\subsection{Filter Selection}
The observations employed two imaging filters for camera 3, F110W and F160W centered at 1.1 and $1.6 \micron$ respectively.  By careful design, the F160W spans the lowest background spectral region available to NICMOS.  This is the minimum between the scattered zodiacal emission that decreases with wavelength and the thermal emission from the warm HST mirrors that increases with wavelength.  Both of these emissions were lower than expected prior to the HDF observations.  The second filter covers a shorter wavelength over a rather broad bandwidth.  This filter provides a second color between the F814W WFPC2 filter and the NICMOS F160W filter. The addition of this filter provides an important discriminator between high redshift star-forming galaxies which will have blue infrared colors and lower redshift galaxies with large amounts of dust extinction which will have red infrared colors. Although very useful for all objects the extra filter is particularly important for objects detected only in the NICMOS bands. Figure \ref{Thompson.fig1.ps} displays the normalized detectivity for the NICMOS filters.  These plots include all of the color dependent terms of the detectivity, including the detector quantum efficiency.  The rapidly changing indices of refraction for most optical coating materials in the 1 micron region account for the complicated shape of the F110W filter.

\placefigure{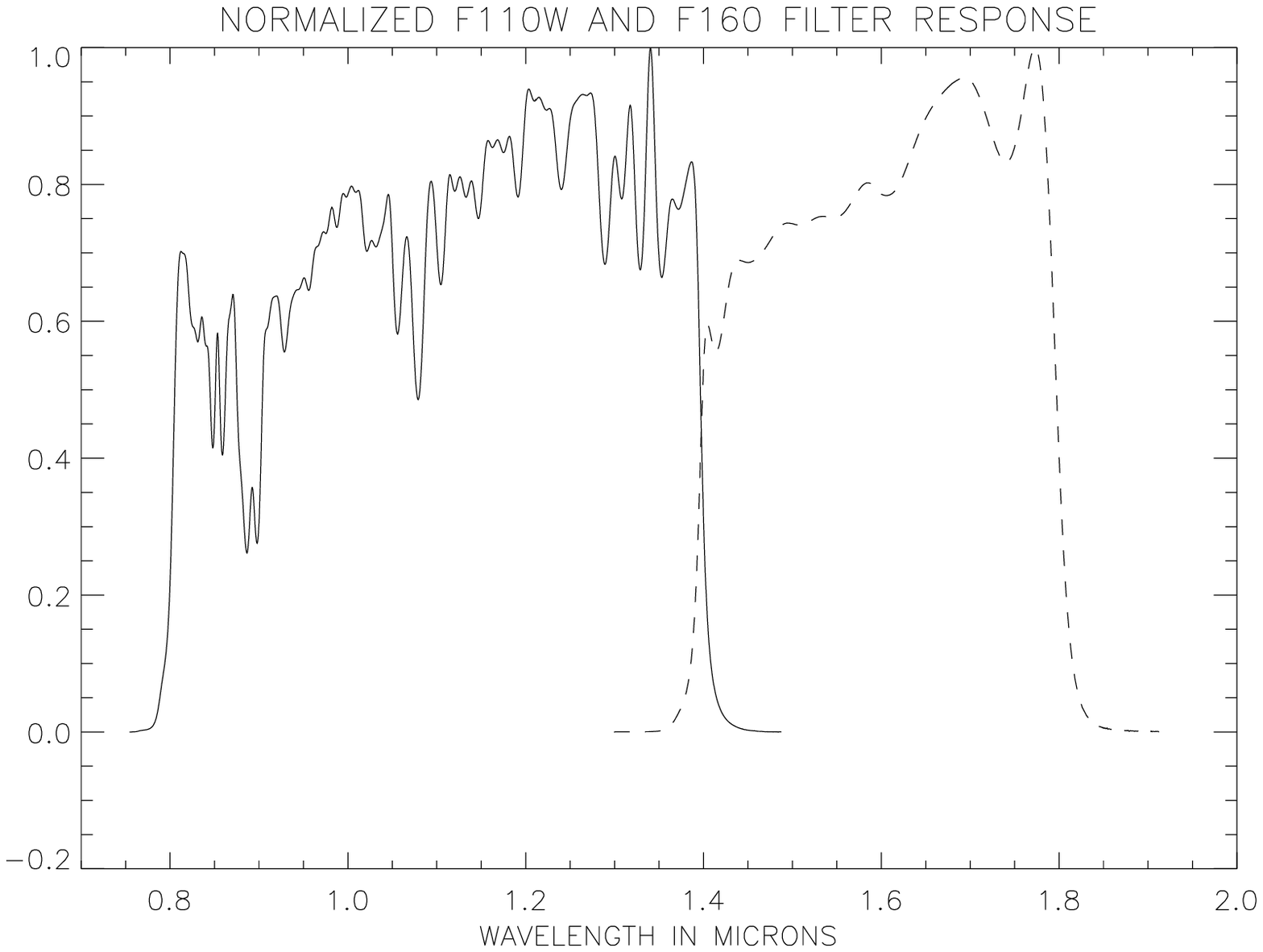}

\section{OBSERVATIONAL STRATEGY}
The preflight decision to devote 49 orbits to each imaging filter and 27 orbits to grism spectroscopy set the parameters for the observing strategy.  The 49 orbits are based on a 7 by 7 grid of positions and the  grism orbits employ 3 different roll angle positions to help remove confusion from overlapping spectra.  There are 3 separate integrations in each orbit to insure that any problems encountered did not necessarily compromise all of the data during the orbit.  Before the observations it appeared that the sky background would be the dominant noise source after about 900 seconds of integration.  The lower than expected sky brightness, however, reduced the sky noise below the read noise for this integration time.

\subsection{Detector Sampling Sequence}
Since there are no bright sources in the HDF the logarithmic NICMOS sampling sequences, designed to handle high dynamic range images, are inappropriate.  Those sequences add amplifier glow noise with several short time integrations near the beginning of the exposures.  Good cosmic ray rejection, on the other hand, requires a sufficient number of samples to establish an accurate signal after any cosmic ray affected samples are rejected.  The HDF integrations utilize the SPARS64 sampling sequence that has evenly spaced 64 second sample times after the first 3 short integration samples.  The integrations have a NSAMP value of 17 which produces a total integration time of 896 seconds.  Three of these integrations fill an orbit.  The total number of integrations in each filter is then 147 integrations of 896 seconds in each filter for a total of $ 1.31712 \times 10^{5}$ seconds or 36.5 hours of observing time per filter.

\subsection{Field Coverage}

Several factors influenced the choice of field size.  The basic purpose of the NICMOS HDF program is deep observations.  This requirement favors very small or no dithers away from the field center.  Accurate background subtraction requires many offsets large enough to insure that most pixels have the majority of their observations off of detectable objects.  Spatial cosmic ray rejection and image resolution enhancement require at least one pixel and fractional pixel offsets respectively.

The pattern of observation positions on the sky is a combination of a small three point dither pattern during each orbit and a larger $7 \times 7$ raster pattern that covered the 49 orbits.  The dither pattern is a 3 position spiral dither with a step size of 0.408 arc seconds, roughly 2 camera 3 pixels.  The x and y spacings of the orbit to orbit raster are 0.918 and 1.523 arc seconds respectively which are 4.5 and 7.5 camera 3 pixels.  The interorbit moves were accomplished with target offsets from the original center position. The basic purpose of the raster was to move the field of view sufficiently that any single pixel had the majority of its integrations with no observable source present.

\subsection{POINTING ACCURACY}

Due to the paucity of bright stars in the HDF region and the roll constraints during the observational time period we were not able to utilize two FGS guide stars.  This situation led to the possibility of roll errors in position about the location of the single guide star.  Real time frequent updates of the gyro bias levels by HST Missions Operations Support Engineering Systems (MOSES) mitigated this problem.  Data provided by the MOSES team (\cite{conr98}) indicated that the positional errors for all orbits used in this paper were less than 0.2 NICMOS camera 3 pixels.  Subsequent analysis discussed in Section \ref{mos} confirmed this data.  Our absolute positions assume that the central star in our field (WFPC2 4-454) has the position stated in the published catalog (\cite{will96}).

\section{DATA REDUCTION}

Data reduction procedures utilized the Interactive Data Language (IDL) software environment for most of the basic data analysis.  KFOCAS (\cite{ads96}), a derivative of FOCAS (\cite{jt81}, \cite{val82})  provided the source detections listed in the catalog.  In order to provide a cross check on the images and catalog presented in this paper we have deliberately reduced and analyzed the data in two separate and independent ways.  Specifically in addition to the IDL and KFOCAS procedures we utilized an independent IRAF based image processing algorithm and an alternative source extraction program, SExtractor (\cite{Ber96}).  The IDL and KFOCAS reduction procedures are described in detail as they produced the bulk of the information on the sources listed in the catalog.  Descriptions of the IRAF and SExtractor reductions are provided when they differ substantially from the IDL and KFOCAS reductions.

\subsection{IDL Image Reduction}
Each 896 second SPARS64 integration produced an individual image.  A set of 55  SPARS64 dark integrations of the same duration as the HDF exposures provided the required dark frames for the analysis.  These dark exposures occurred just prior and coincident with the HDF exposures.  The data reduction procedures produce a completely processed image for each integration.  Section \ref{mos} describes the combination of the images into the final mosaic.

\subsubsection{Dark Frames}

Dark frame reductions begin with the division of each 17 sample SPARS64 darks into 16 first differences.  A first difference is simply the difference between a readout and the previous readout.  The first differences are then combined via a sigma clipping mean to produce a final super frame that is free of cosmic rays to the 3 sigma level of the 55 combined observations. Although for most observations a simple median of first differences would suffice, observations at the sensitivity level of the HDF required sigma clipped means to avoid digitization noise.  The average camera 3 dark current is 0.2 electrons per second.  In each 64 second first difference about 12.8 electrons accumulate.  The detector gain for camera 3 is 6.5 electrons per ADU for an average of 2 ADUs per first difference.  If medians are taken of these observations there would still be 50\% noise even for an infinite number of integrations.  The first differences are then recombined to produce a ramp dark for subtraction from the imaging integrations.  The ramp dark is a sequence of summations ranging from just the initial first difference, the initial and second first difference, to the total of all of the first differences.

\subsubsection{Image Frames}

Analysis of the image integrations starts with the production of a set of 16 ramp readouts for the 17 samples from each SPARS64 integration.  Subtraction of the super dark ramp from each integration produces a set of dark current corrected but not sky corrected integrations.  A set of standard linearity corrections are next applied to pixels that have exceeded the linear signal response region but have not saturated.  In practice only bad ``hot" pixels  receive a correction due to the low signal levels in the HDF observations.

After correction for linearity the integrations are corrected for cosmic rays by fitting a linear function to the ramp values.  The slope of this function is the signal rate in ADU per second.  Cosmic rays produce an instantaneous discontinuity in the signal function.  Subtraction of the fitted function from  the signal produces an output that has a distinctive S shape if a cosmic ray is present.  In one readout the difference between the fit and the signal transitions from negative to positive.  Detection of a transition greater than expected from noise indicates the presence of a cosmic ray.  The offending first difference is then removed from the ramp and a new fit is calculated.  The new fit is again checked for cosmic rays and detected cosmic rays are removed in a similar matter.  Any fits still beyond the expected noise are declared bad and flagged as bad.  If the cosmic ray produces saturation, only the readouts before saturation are used in the final fit.  All detected cosmic rays are recorded in the data quality image extension.  This procedure is unique to the NICMOS instrument on HST due to the ability to nondestructively read out the detector during the integration.  Further cosmic ray removal can occur if necessary during the image mosaic construction in the standard manner.

Before the flat field can be applied all of the quadrant biases in the individual images must be removed.  If there is a bias level in the image, the flat field function will produce variations in the bias which will remain in the data.  The removal procedure can be iterative but in practice one iteration is sufficient.  The first step produces a median of all of the cosmic ray corrected images.  This median is then subtracted from each individual image.  If there are no quadrant biases the median value in each detector quadrant should be zero since the dark current and sky are removed by the subtraction.  The sources do not dominate the image so they contribute very little to the median.  The second step measures the median of each quadrant in each image and records the value.  The final step then subtracts these quadrant bias levels from each cosmic ray corrected image.  The flat field correction was then applied to each image.

The bad pixels are marked from a bad pixel mask determined from the previous observing history with the NICMOS camera 3.  The bad pixel mask contains both low response pixels and hot pixels defined as pixels with an excessive dark current.  In each image the bad pixels were replaced by the median of the total image.  The drizzle mosaic process does not use pixels marked as bad in the final image construction.

The median of all of the final individual images determines the level of the sky emission.  The raster and dither of the large number of images reduces the source contribution of the median to a value less than the expected noise level.  Inspection of the median image did not reveal any source contributions.  Subtraction of this sky level from each image completes the analysis of the individual images.  The median sky level in the F160W filter is 0.55 electrons per second which is lower than the original estimates prior to the observations.  This is not surprising as one of the selection criteria for the HDF was a low zodiacal background.

\subsection{IRAF Image Reduction}

For comparison with the IDL procedures images were reduced in an independent pipeline using NICRED 1.5 (\cite{mcl97}, \cite{leh98}) and modified IRAF scripts developed to reduce Camera 3 images taken in parallel mode (\cite{yan98}). Two median sky-dark frames were produced, one from the first exposure in each orbit and one from combining the second and third exposures in each orbit, to minimize the effect of  any pedestal in the first exposure. These were used as dark frames along with the same flats utilized in the IDL reductions.  These flats, observed on Dec. 23, 1997 are identical to the flats used in the IDL reduction.   The residual bias levels in the individual quadrants were removed by fitting a Gaussian to a histogram of the pixels in each quadrant, and subtracting the peak value.  A new bad pixel mask was created from the exposures. The images were inspected and any remaining cosmic ray hits or satellite trails were individually masked.

\subsection{Mosaic Techniques}
\label{mos}

Both of the data reduction procedures utilize modified versions of the drizzle software developed for the reduction of the WFPC2 HDF images.  The drizzled pixel size in each case is $\sim 0.1 \arcsec$, one half of the original NICMOS Camera 3 pixel size.  The drizzle parameter PIXFRACT is 0.6 in the drizzling of the IDL reductions while it was set to 0.65 in the drizzling of the IRAF results.

\subsubsection{IDL Image Mosaic}
 
The first task of mosaic production is an accurate determination of the relative offsets between the individual integrations.  We compared offset information from the world coordinates in the header files, shifts computed from the IRAF/STSDAS Dither package, offsets from the IRAF imcentroid package of five individual objects in the field and finally individual inspection via interactive IDL tools.  The NICMOS geometric distortions have been determined to be negligible so no geometric distortion corrections were made.
  
In general, the agreement among the four methods was quite good. For the F160W images the discrepancy between the world coordinate shifts and the IRAF generated shifts fell between -0.2 and 0.4 pixels in x with a mean of 0.15 (RMS=0.03) and a range of -0.5 to 0.2 pixels in y with a mean of -0.15 (RMS=0.03).  The internal difference between the shifts determined by IRAF procedures averaged -0.02 pixels.

For the F110W filter the difference between the world coordinate shifts and those determined by IRAF procedures varied from -2.0 to 0.7 pixels in x and
-0.4 and 2.7 pixels in y with means of 0.1 (RMS=0.14) and 0.7 (RMS=0.36)
respectively.  The mean internal difference between the IRAF procedures was -0.12 for x and 0.04 in y.  The large excursions of 2.0 and 2.7 pixels in x and y were seen in only two images.  The IRAF Dither and imcentroid positions agreed to 0.1 pixels in these images.  Visual inspection of the images confirmed the IRAF positions. In both the F160W and the F110W filter the rotation angle varied by less than 0.005 degrees due to the excellent effort of the MOSES group in limiting the roll during the single guide star observations.

The data were drizzled using Drizzle Version 1.2 February, 1998 (\cite{fru97}) with image offsets derived from the mean of the IRAF procedures since in cases where the IRAF positions differed from the world coordinates interactive inspection via the IDL tools confirmed the IRAF positions. As discussed above no geometric distortion correction or image rotation was required. High cosmic ray persistence noise levels after transit of the Southern Atlantic Anomaly required removal of 28 F160W integrations and 36 F110W integrations from the final mosaic image.  A comparison of a straight combination of the drizzled frames and a combination averaged with $3\sigma$ clipping showed no differences indicating that the IDL cosmic ray removal techniques were effective.

\subsubsection{IRAF Image Mosaic}

The IRAF reduction images were drizzled (Drizzle Version 1.1, Fruchter
et al.~1997)  with offsets determined from the centroid of the central 
star (NICMOS 249) in each frame.  No rotation or geometric distortion 
corrections were necessary. Due to persistence of cosmic rays encountered in 
SAA passages, 25 F160W images and 37 F110W images were removed from the final 
mosaic. Which frames to remove was determined independently by inspection 
which leads to the slight difference from the number not included in the IDL 
image mosaic.  The drizzled frames were averaged with 3$\sigma$ clipping 
to remove any residual low level cosmic rays. 

\section{THE IMAGES}

Figures \ref{Thompson.fig2.ps} and \ref{Thompson.fig3.ps} show the F110W and F160W images respectively, produced by the IDL image reduction and drizzle procedures described in the preceding section.  The raster pattern of observations produces much lower signal to noise areas in the image at the edges where the number of overlapping integrations are greatly reduced. These areas are deleted from the image even though many strong sources are evident in these regions.  The area covered by the images is a $49.19 \times 48.53 \arcsec$ ($481 \times 476$ pixels) rectangle.  Figure \ref{Thompson.fig4.ps} is a color composite of the two infrared images and the F606W WFPC2 image.  The WFPC2 image has been rotated and resampled to fit the orientation and pixel size of the NICMOS images.  The red, green, and blue colors represent the F160W, F110W, and F606W intensities.  As with the original WFPC2 color image, the stretch and color curves have been manipulated to show faint objects while preserving the detail of features in the brighter objects.  This image should not be used for quantitative purposes.  Figures \ref{Thompson.fig2.ps} and \ref{Thompson.fig3.ps} are also stretched to show the best range of features.  The very high dynamic range of the image can not be displayed in a linear intensity image.

\section{SOURCE DETECTION AND PHOTOMETRY}

Since the original WFPC2 HDF catalog (\cite{will96}) utilized KFOCAS to generate its listings our primary catalog listings also utilize KFOCAS to provide consistency.  We also provide a description of the SExtractor source extraction process.  The main difference between KFOCAS and FOCAS is the utilization of a supplemental image  by KFOCAS that specifies the relative detectivity at each point in the image.  This is important for the NICMOS HDF images where there are significant variations of quantum efficiency and total integration time over the image area.

\subsection{Estimation of the input sigma for KFOCAS}
\label{sig}

KFOCAS uses a constant $1\sigma$ level that is either determined from the first few lines of the image or is input manually by the user.  Since the first few lines of the NICMOS image have much lower signal to noise than rest of the image we estimated the $1\sigma$ value manually from a histogram of the signal levels in all of the pixels.  Figure \ref{Thompson.fig5.ps} gives the histogram of the pixel values in the two dithered images shown in figures \ref{Thompson.fig2.ps} and \ref{Thompson.fig3.ps}.  Only the pixels from the area covered by the images are used in this histogram.  The histogram peaks at zero signal as expected for sky subtracted observations.  The long extensions of the histograms toward positive values due to the sources in the field are cut off in this figure.  The flat fielding process multiplies the true noise value in the image by the value of the flat field.  This process raises the noise level in low quantum efficiency areas and lowers it in high efficiency areas.  Since the median efficiency of the area is set to one this process should not appreciably alter the width of the curve.

\placefigure{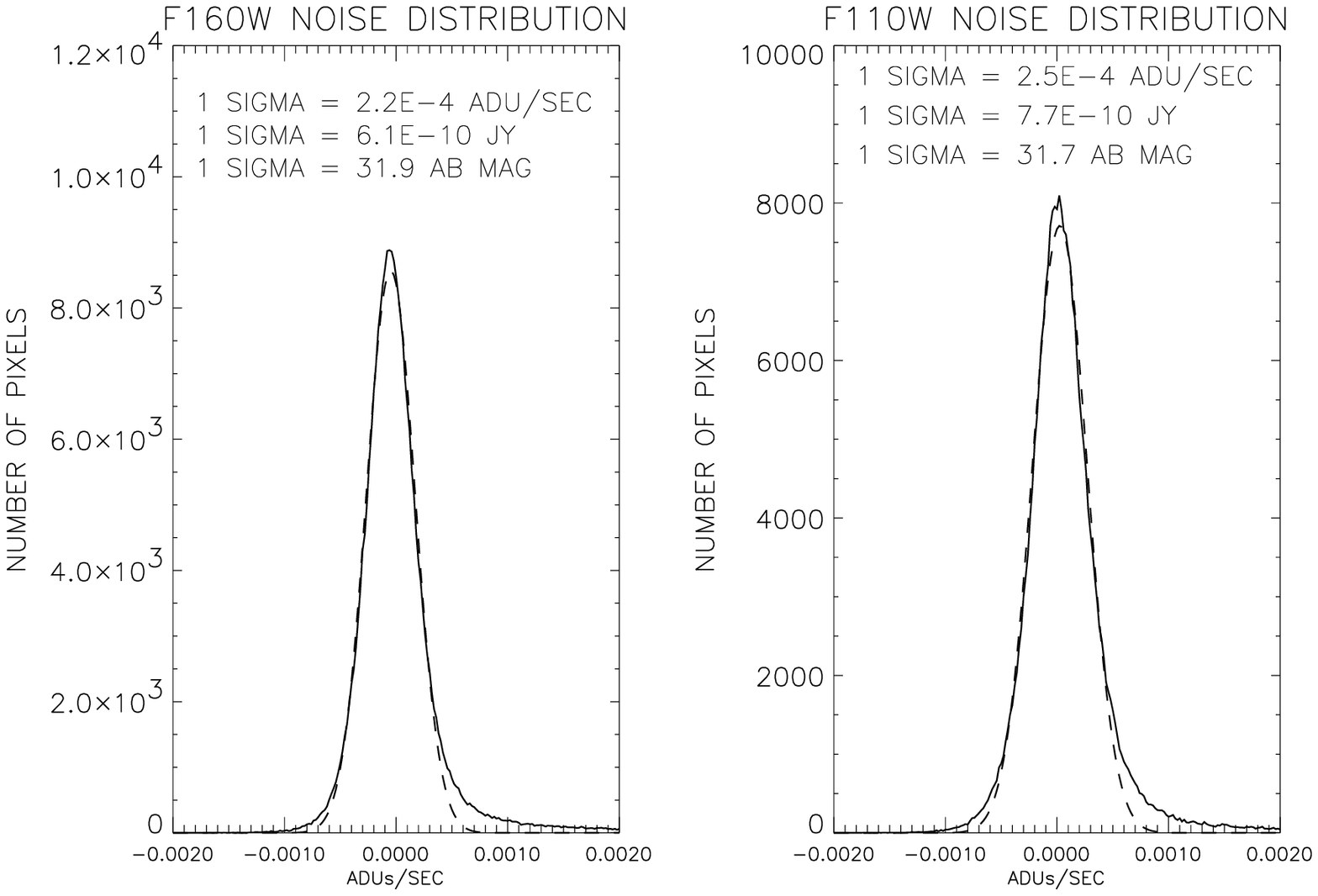}

As learned in the production of the WFPC2 images, the drizzling process produces a correlated image and hence correlated noise (\cite{will96}, \cite{fru98}).  There is approximately a factor of two reduction in the apparent noise as a result of the drizzle process for a factor of two reduction in linear pixel size.  The numbers given in figure 5 should therefore be multiplied by a factor of 2 to determine the true $1\sigma$ value of the noise.  This gives the noise figures of $1.22 \times 10^{-9}$ Jy for the F160W filter and $1.54 \times 10^{-9}$ Jy for the F110W filters.  These are the powers that produce a signal equal to a $1\sigma$ noise in a single pixel.  Use of these levels resulted in KFOCAS missing a large number of real sources easily identified by eye.  We therefore dropped the $1\sigma$ estimates to a very low value of $5.5 \times 10^{-10}$ Jy or $2.0 \times 10^{-4}$ ADUs per second for the F160W filter and $3.5 \times 10^{-10}$ Jy or $2.3 \times 10^{-4}$ ADUs per second for the F110W filter.  The number of $\sigma$ for the detection limit was then varied until all known real sources were detected without excessive over selection. It is of course the product of the chosen $1\sigma$ noise level and the number of sigma for detection parameter that determines the signal value a pixel must have to be considered a potential source. A known real source is an object easily seen by eye in the NICMOS images that is exactly coincident with an observed source in the WFPC2 HDF images.  A more rigorous discussion of the completeness and reliability of the selected sources occurs in sections \ref{comp} and \ref{rel}.  The catalog listings are limited to sources with signal to noise ratios that exceed or equal 2.5.  This discards some sources that by many tests appear to be real but eliminates a large number of sources that have a significant chance of being false.

\subsection{KFOCAS Reduction}

We prepared the drizzled images for the KFOCAS procedure by multiplying the signal in ADUs per second by $10^{5}$ and subtracting the minimum value, a negative number, from the multiplied image.  This produced an image that had no negative values and where all of the significant values were well represented in the integer arithmetic used by KFOCAS.  The zero point magnitudes of the modified images are 35.3 for the F160W image and 35.186 for the F110W image.  The source extraction utilized the standard KFOCAS procedures of the series KDETECT,  KSKY, KEVALUATE, KSPLIT and RESOLUTION.  Our drizzled pixels have 6.25 times the area of the WFPC2 drizzled pixels, therefore, we set the minimum area for detection in pixels to 2, to avoid missing very compact galaxies.  The parameters for the KFOCAS reduction are listed in Table \ref{kfoc}.  The point spread function (PSF) matrix for smoothing the data is a $3 \times 3$ matrix that mimics the PSF of the central star in the drizzled data.  This is much more sharply peaked than the Gaussian function used in the SExtractor analysis.

\subsubsection{Preparation of the detectivity image}
\label{qualim}

Each pixel, $pix_{i,j}$, in the final image has a quality $Q_{i,j}$ value associated with it.  The quality value is the square root of the sum of the squares of the total efficiency of each pixel in the individual image that contributes to the final image.  Due to the raster and dither pattern a pixel in the final image has contributions from many different individual image pixels.
\begin{equation}
Q_{i,j} = \sqrt{\sum_{k=1}^n (eff_{k})^2}
\end{equation}
Here n is the number of pixels contributing to the final image pixel, $pix_{i,j}$, and the efficiency $eff_{k}$ of each contributing pixel is measured by the inverse of its multiplicative flat field value. Figures \ref{Thompson.fig6.ps} and \ref{Thompson.fig7.ps} show the detectivity functions over the image areas used in the catalog.

\placefigure{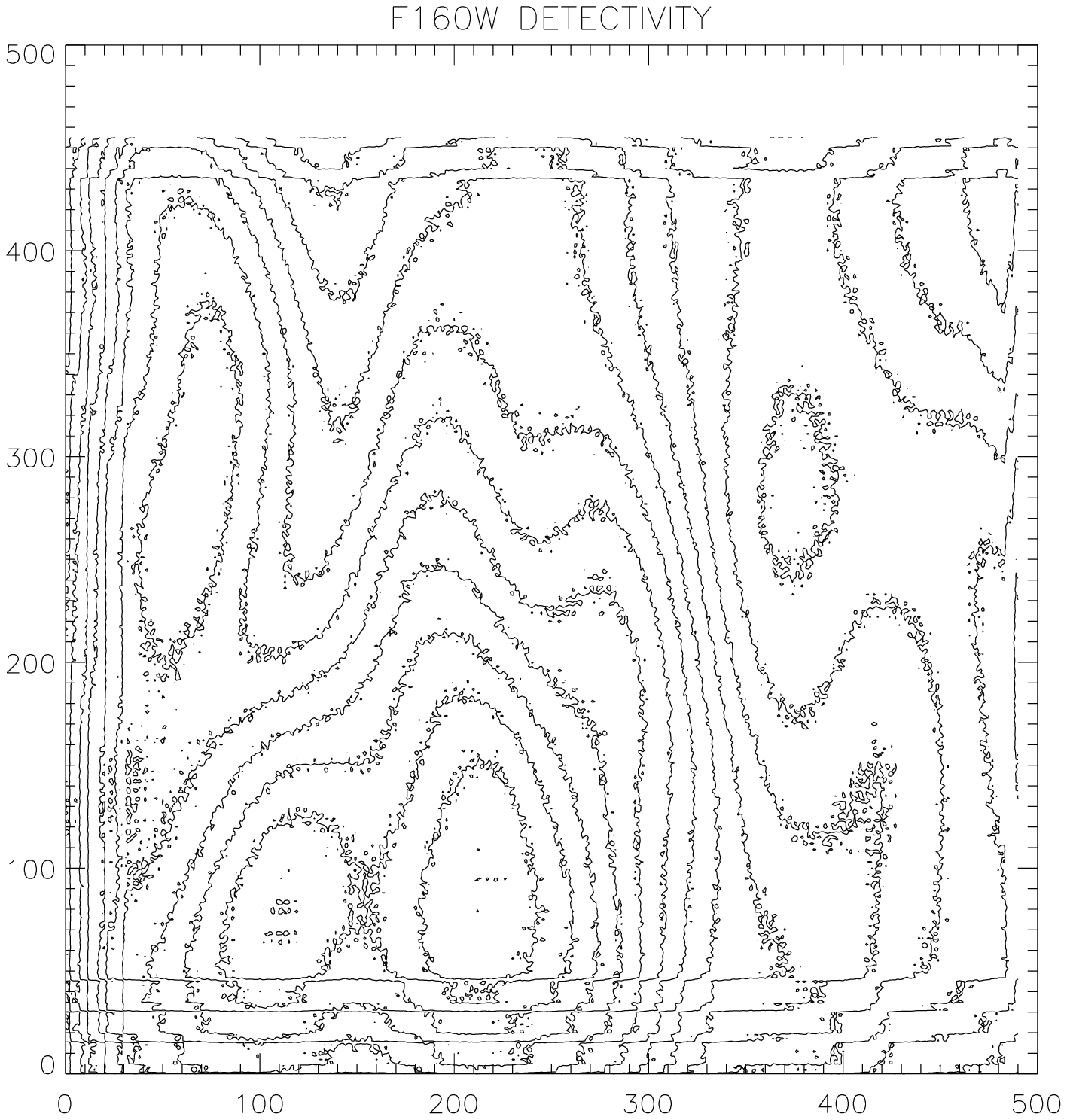}
\placefigure{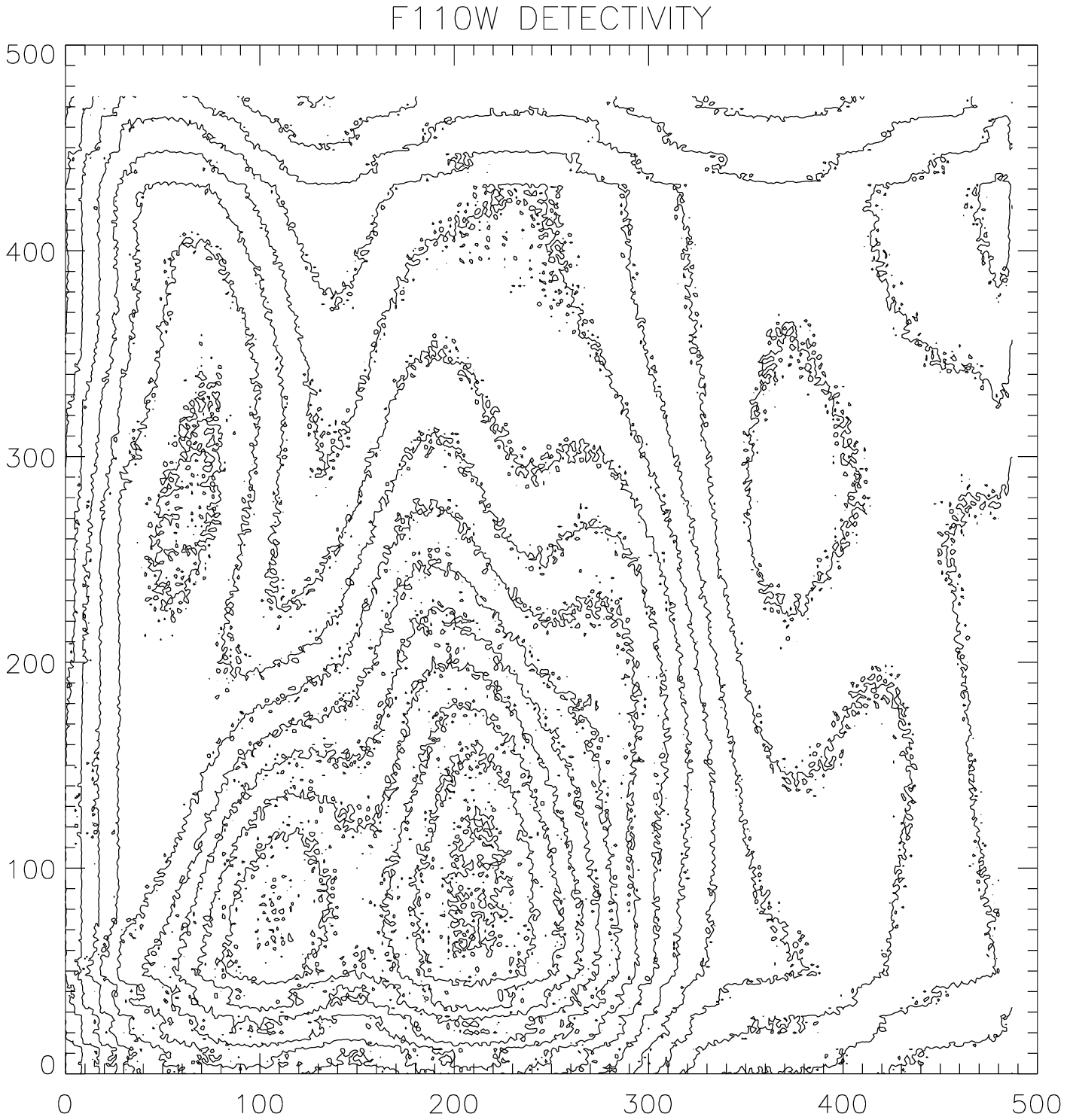}

\subsection{SExtractor Reduction}

As a cross check on the KFOCAS detections we utilized an alternative source extraction system, SExtractor, on data reduced via the IRAF reduction procedure rather than the IDL based procedure. 

\subsubsection{Galaxy Detection}

We performed object detection on the F160W images and photometry on 
the F160W and F110W images using SExtractor version 2.0.7 (\cite{Ber96}).
The final F160W and F110W images reduced in the IRAF pipeline showed low frequency
structure in the background in the X-direction. We  created
background frames in SExtractor using a $64 \times 32$ mesh which were
subtracted from the reduced frames,
producing cosmetically more uniform backgrounds.
We measure $1\sigma$ noise levels from histograms of all the pixels in the frames
of $2.0 \times 10^{-4}$ ADUs per second in the F160W image and $2.4 \times
10^{-4}$ ADUs per second in the F110W frame, consistent with the values
shown in Fig. \ref{Thompson.fig5.ps}.
The amplitude of the fluctuations in the background varies by up to 50 \% across the
image due to variations in the quantum efficiency of the detector and the dither
pattern used. Thus we used the option in SExtractor (WEIGHT\_TYPE) which accepts
a user supplied variance map (for which we used the `detectivity' function described
in Section \ref{qualim}).  SExtractor robustly scales
the weight map to the appropriate absolute level by comparing the weight map
to an internal, low resolution, absolute variance map built from the science
image itself. In contrast to the
KFOCAS source extraction, all object detection was done on the F160W image and
magnitudes were measured to the corresponding isophotes on the F110W image.

After experimenting with different values of SExtractor parameters, we adopted the
values given in Table \ref{t_sextparam}. Aside from the determination of the local variance
itself, the three most critical parameters which affect the detection of
very faint isolated sources are FILTER\_NAME, DETECT\_MINAREA and DETECT\_THRESH. The
FILTER\_NAME parameter describes the smoothing kernel which is applied to the image
and for this a Gaussian with a full width half-maximum of 2.0 (drizzled) pixels
was used over a 3 x 3 pixel grid. 
We tested various combinations of DETECT\_MINAREA, the minimum number of contiguous 
pixels above a level which is the product of the parameter DETECT\_THRESH times the local RMS fluctuation in the background.  Our final choices for these parameters are DETECT\_MINAREA  = 2 pixels and DETECT\_THRESH = 2.15$\sigma$. This choice  for DETECT\_MINAREA (the
same value as for the equivalent KFOCAS parameter) favors slightly the
detection of the most compact sources and the final choice of 2.15 
for DETECT\_THRESH was dictated by an attempt to strike a judicious balance 
between completeness and reliability. We tested an alternative set of parameters with DETECT\_MINAREA set to
3 pixels and  with DETECT\_THRESH to a lower value in order to detect the same
number of sources as with DETECT\_MINAREA =2 and
DETECT\_THRESH=2.15.  Estimating the number of false detection rates and
completeness as discussed in sections \ref{comp} and \ref{rel}, we found that these two sets of parameters behaved very similarly. For the reason stated
above, we selected  DETECT\_MINAREA=2. Sections \ref{comp} and \ref{rel} contain a  more quantitative discussion of the completeness and reliability of the detected sources.

Although like the KFOCAS reductions we have deliberately erred on the side of extracting a fairly large estimated fraction of false detections  (at the faintest levels) and a completeness level which is only of order 50\%, the catalog listings only contains sources with signal to noise ratios greater than 2.5. 

\subsection{Comparison of the IDL-KFOCAS and the IRAF-SExtractor Photometry}
\label{kscomp}
At this point our analysis contains four components, the IDL and IRAF reduced images and the KFOCAS and SExtractor source extractions and photometry.  In the spirit of independent cross checks it is useful to compare these results and to see if any differences lie primarily in the images or in the source extraction procedures.  We will discuss the differences in completeness and reliability between the two methods after sections \ref{comp} and \ref{rel}.  Data presented so far have been for either the IDL-KFOCAS or IRAF-SExtractor procedures. The third panel in fig. \ref{Thompson.fig8.ps} shows a comparison between the aperture magnitudes found by KFOCAS in the IDL image and the aperture magnitudes found by SExtractor in the IRAF images for all objects detected in common.  The delta magnitudes are KFOCAS magnitude minus SExtractor magnitude.  In both cases the diameter of the aperture is 0.6\arcsec.

\placefigure{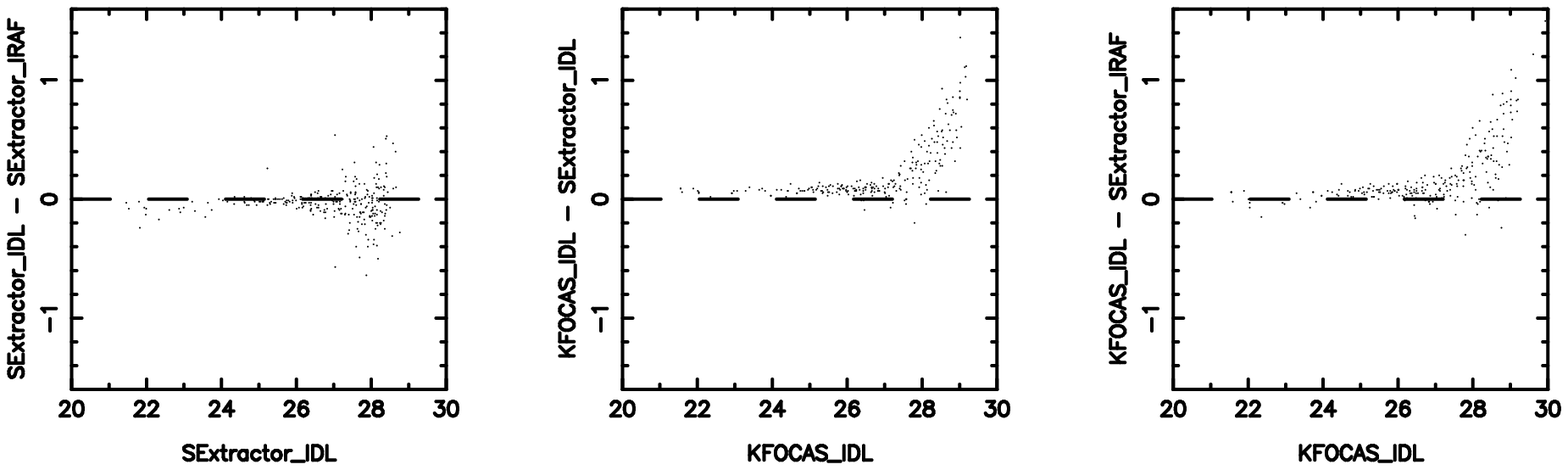}

As expected the correlation is very good at bright magnitudes and gets worse at the faint end.  The SExtractor magnitudes are equal to the KFOCAS magnitudes up until 28th magnitude where the SExtractor magnitudes become significantly brighter than the KFOCAS magnitudes. These differences could be either to differences in the input images or in the magnitudes extracted by KFOCAS and SExtractor. To check this we ran SExtractor on both the IDL reduced and the IRAF reduced images.  The first panel in fig. \ref{Thompson.fig8.ps} shows the result of this test which shows a uniform scatter about
the zero level at the fainter magnitudes.  The slight offset at the bright end is probably due to the IRAF procedure IMCOMBINE clipping some pixels in the brightest galaxies in the IRAF image.  A follow up test  comparing the KFOCAS reductions of the IDL image with SExtractor reductions of the IDL images is shown in the middle panel of fig. \ref{Thompson.fig8.ps}. This plot is essentially identical to the last panel except for about a 0.07 magnitude offset at the brighter end.  This set of tests shows that the differences between the KFOCAS/IDL magnitudes and the SExtractor/IRAF magnitudes shown in  the last panel of fig. \ref{Thompson.fig8.ps} are entirely due to the differences between the KFOCAS and SExtractor algorithms, not from any differences between the IDL and IRAF image production procedures.  The origin of this difference is not clear but the reader should be aware that these two standard procedures do produce differences at the very faintest levels.

\section{THE CATALOG}
\label{cat-ref}

Table \ref{cat} contains the catalog of sources from the KFOCAS source extraction from the F160W and F110W images. This catalog only contains sources with signal to noise ratios greater than 2.5.  We anticipate a future publication describing the sources with less reliable detections. The catalog contains  342 objects, some of which are components of a larger object. The catalog contains 235 objects in common with the WFPC2 catalog. 221 objects have detections in both filters, 56 objects have only a detection in the F160W filter, 53 have only a detection in the F110W filter, and none have detections only in SExtractor.  The objects are arranged by right ascension which sometimes separates different components of the same object in the catalog.  The data and numbering in the catalog have the priorities, in order, of F160W KFOCAS, F110W KFOCAS, and SExtractor. This means that all KFOCAS objects detected in both filters use the KFOCAS F160W RA, DEC, x, and y positions.  The magnitudes come from the KFOCAS F160W and F110W extractions.  Positional coincidence is relative to the F160W positions.  Objects that have F110W KFOCAS detections but not KFOCAS F160W detections use the F110W KFOCAS positions and magnitudes.  The catalog columns contain the following parameters.

\textbf{ID:}  This is a running number for each object.  The numbers after the decimal point indicate the level of splitting by KFOCAS up to three levels of daughter objects.  Since the list is arranged by right ascension daughter objects can appear separately from the parent objects.  No object is repeated.  Numbers of 900 or higher are split F110W objects that are not coincident with any F160W split even though some of the components are in common.

\textbf{WFPC:} The WFPC column lists the nearest WFPC2 source from the \cite{will96} catalog.  This is not necessarily the same object, just the nearest.

\textbf{s:}  The s value is the separation in arc seconds between the NICMOS and the nearest WFPC2 object as listed in the WFPC column.  A large value of separation indicates that the NICMOS and WFPC2 object are probably not associated.

\textbf{x,y:} These columns give the x and y values of the centroid of the source in the F160W or F110W image.  If the object is detected in both images the x and y values refer to the F160W image. Objects detected only by SExtractor have the values determined by SExtractor.  This order of precedence holds for all of the subsequent values.

\textbf{RA,DEC:}  These columns give the right ascension and declination of the centroid of the source.  Only the minutes and seconds are listed.  The hour of right ascension is 12 hours for all sources and all sources have 62 degrees of declination.  The source positions assume that the central star, NICMOS 145 and the WFPC2 object 4-454 have the same position and that the measured plate scales of the NICMOS camera 3 are correct.  In this sense all positions are relative to the position of the WFPC2 4-454 object.  

$\mathbf{t_{160}}$,$\mathbf{i_{160}}$,$\mathbf{a_{160}}$,$\mathbf{t_{110}}$,$\mathbf{i_{110}}$,$\mathbf{a_{110}}$:  These are the  total, isophotal and aperture magnitudes found by KFOCAS in the F160W and F110W images. The aperture diameter for the aperture magnitudes is $0.6\arcsec$.  The total and isophotal magnitudes are as described in \cite{will96}. A value of 99.99 indicates that the object was not detected in that filter.  The F160W and F110W objects are considered to be in common if they lie within $0.25\arcsec$ of each other.  If the last digit in the tests parameter (see below) is 3, the magnitudes are from the SExtractor procedure.

\textbf{S/N:} The signal to noise value quoted in the catalog is calculated by the same technique used in the optical HDF.  The value of the signal to noise is $S/N = L_{i} / \sigma (L_{i})$ where the variance $[\Gamma \sigma (L_{i})]^{2}$ is,
\begin{equation}
[\Gamma \sigma(L_{i})]^{2} = \Gamma N_{obj} + 1.9^{2} \Gamma^{2} \sigma_{sky}^{2} A_{obj} + 1.9^{2} \Gamma^{2} \sigma_{sky}^{2} A_{obj}^{2} / A_{sky}
\end{equation}
as quoted in its correct form by (\cite{poz98}).  For NICMOS the value of $\Gamma$ in electrons per ADU is 6.5. $L_{i}$ is the sky subtracted number of counts, $\sigma_{sky}$ is the sky sigma in ADUs, and the object and sky areas $A_{sky}$ and $A_{obj}$ are the areas in pixels returned by KFOCAS.  This equation reformulated in terms of count rates in ADUs per second is given by
\begin{equation}
\sigma_{tot}^{2} = rate_{obj} / (\Gamma t) + 1.9^{2} \sigma_{sky rate}^{2} A_{obj} + 1.9^2 \sigma_{sky rate}^{2} A_{obj}^{2}/A_{sky}
\end{equation}
Here  $rate_{obj}$ is the sky subtracted source count rate in ADUs/sec, $\sigma_{sky rate}$ is the sky sigma value in ADUs/sec, and t is the integration time in seconds.  The final signal to noise is $rate_{obj} / \sigma_{tot}$.  The measured value of the sky sigma is $2.2 \times 10^{-4}$ ADUs per second for the F160W image and $2.5 \times 10^{-4}$ for F110W as shown in Fig. 2. The factor of 1.9 in each of the equations is the estimated value of the noise correlation discussed earlier.  Figure \ref{Thompson.fig9.ps} shows a plot of the signal to noise ratios calculated by this method for the KFOCAS determined sources.

\textbf{A:}  This is the isophotal area of the source in square arcseconds as determined from the value returned by KFOCAS.

$\mathbf{r_{1}}$:  This is the half light radius returned by KFOCAS.

\textbf{tests:}  This parameter indicates which of the various reliability tests the source passed.  A source that passed all tests has a value of 22111, one that passed no tests has a value of 00000.  The first number is 0,1,or 2 if the source was detected in none, one, or both of the F160W half catalogs. For an explanation of the half catalogs see Section \ref{half}. The second number is the same test in the F110W catalog.  The third number is 0 or 1 depending on whether the source was found in both NICMOS bands of the full image extractions. The fourth number is 0 or 1 depending on whether the source is detected in the WFPC2 catalog.  The last number is 0 or 1 depending on whether the source was found in the independent Sextractor catalog. Note the discussions in sections \ref{krel} and \ref{srel} on the differences in half catalog detection probabilities between KFOCAS and SExtractor.  In all cases a common detection means that the source centroids lie within 0.25 arc seconds of each other.  No color or magnitude tests are applied as part of the common object association.

\placefigure{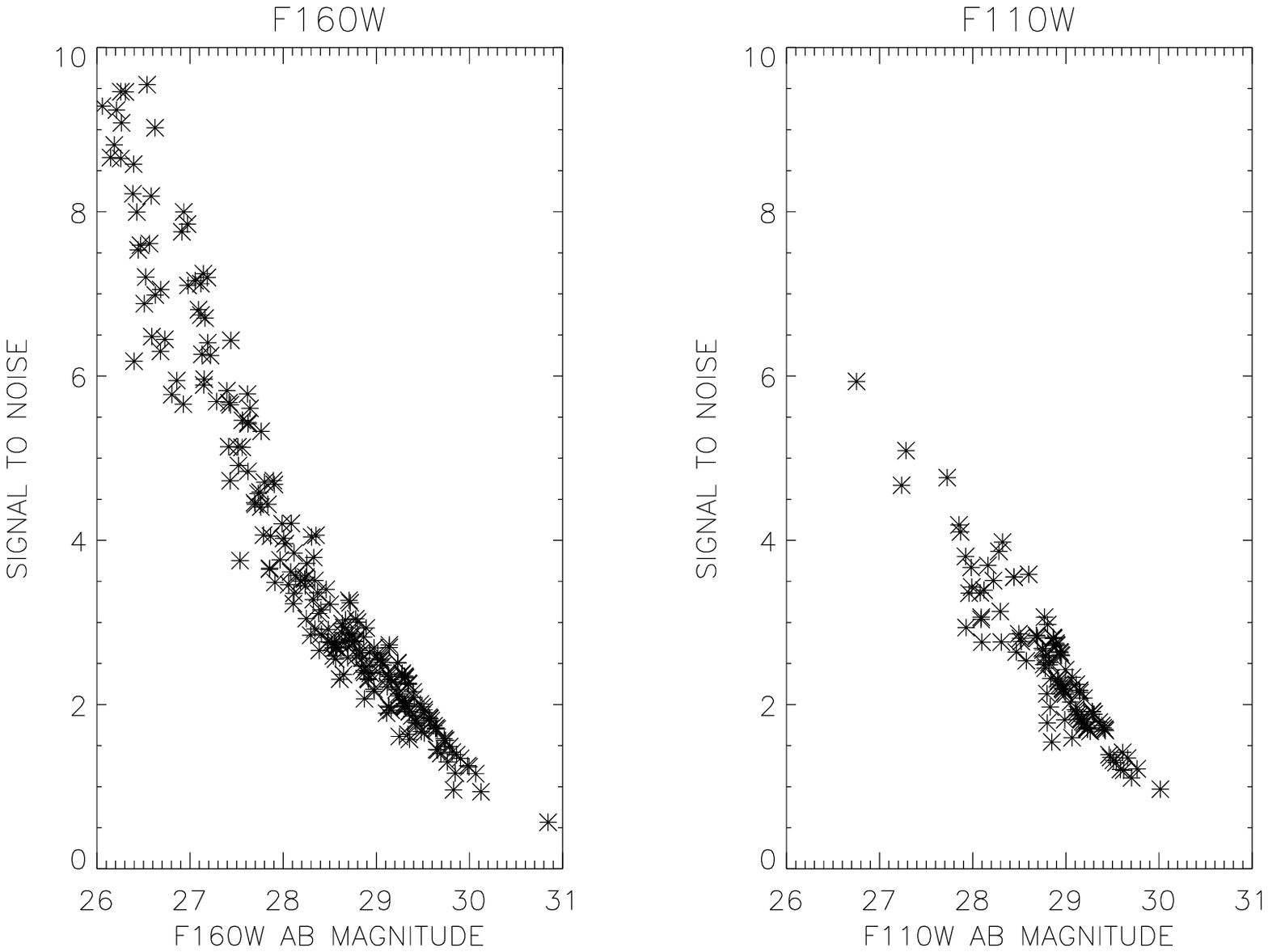}

\section{Completeness}
\label{comp}

Our calculated $50\%$ completeness levels are KFOCAS AB aperture magnitudes of 28.7 and 28.2 for point objects and extended sources in the F160W filter and KFOCAS AB magnitudes of 28.6 and 28.0 respectively in the F110W filter.  These limits are based on a technique of adding sources to the image at various flux levels and running the source extraction programs to see what percentage of the added sources are recovered. These numbers are based on the KFOCAS reductions. 

\subsection{KFOCAS Completeness}
\label{kcomp}

The test for the KFOCAS program established a regular grid of 49 positions in a 7 by 7 pattern that evenly covered the image area utilized in the catalog extraction.  Sources are placed at these positions to create a new image which contains the original image plus the added sources.  KFOCAS then creates a new catalog utilizing exactly the same parameters used in the final catalog preparation.  An automatic program checks the new catalog to see what percentage of the added sources are recovered by KFOCAS.  The sources are sequentially dimmed in half magnitude steps from their original magnitudes of about 21 to the final magnitude of 32.  The added sources are real sources extracted from the final image.  The NICMOS source 145, the central star, represents a point object and the elliptical galaxy, NICMOS 212, is the extended object.  These are WFPC2 sources $4-454$ and $4-471$ respectively. Alteration of the position of the grid confirmed that the completeness limits were consistent at different grid positions.  The extended source completeness limit will of course be brighter for more extended objects than NICMOS 212 but this galaxy is one of the largest in the image.

The completeness value at a given magnitude is for the magnitude of the input source not the magnitude at which the source is recovered.  This value is approximately equal to the recovered magnitude for sources brighter than 26.5.  At the very faintest levels the recovered magnitudes are 0.3 to 0.5 magnitudes fainter than the input magnitudes.  The measured completeness limits are fit by 
\begin{equation}
(1 - {\frac{m - m_{b}}{m_{a} - m_{b}}})^n \sqrt{\frac{m_{b}}{m}}
\end{equation}

In this function $m_{a}$ is the magnitude at which the completeness goes to zero, $m_{b}$ is the magnitude where the completeness is $100\%$ and the power index n is adjusted to fit the data.  The function is purely empirical, simply designed to fit the data well.  This function smooths the curves and provides the interpolation from the observed magnitudes to the magnitudes used in table \ref{cr}.  As expected from the nonuniform quality map of the image, the completeness limit is not uniform over the image.  The completeness is somewhat lower in the right hand half of the image.  The completeness limits in table \ref{cr} should be considered as an average across the image.

\subsection{SExtractor completeness}

For the SExtractor catalog we used an analysis similar to the one described 
in Yan et al.~(1998).  We selected a compact galaxy (4-289 in
the Williams catalog),
representative of the majority of the objects in our field, and dimmed it
in half-magnitude increments from 25.5 to 29.
The dimmed galaxy was dropped at random
positions 10,000 times, superposed on the full image and two images
with half the exposure time of the full image.  
(The ``half'' images are discussed in section \ref{half}.)
We ran SExtractor at the position where the galaxy was dropped at
each iteration to determine whether or not the dimmed galaxy was recovered
and at what magnitude.
The use of random positions in the simulations
allows us to include completeness corrections arising from
non-detections and magnitude errors caused by crowding and 
spatially dependent errors in the sky subtraction and flat field correction.
As discussed in section \ref{kcomp}, the
completeness values for recovery of the images are again an
area-average of the completeness
since the sensitivity across the NICMOS images is not uniform.
These same experiments also give us the matrix relating the input 
and output aperture magnitudes for those
galaxies which are recovered.
The input magnitude and mean recovered magnitude agrees to 
within 0.07 magnitudes in the bins through 28. At 
28.5 the mean of the recovered galaxy magnitudes begins to 
brighten (28.21) and by an input magnitude of 29 it 
brightens substantially (28.11). 
The galaxies that land on negatives noise regions
are lost completely. 
The SExtractor catalog becomes 50\% complete at AB$\approx$28.3
for the compact galaxies in our survey.
 
To test differences in detectability of the point spread function
due to the source landing at different locations within the 0.2"
pixels, we ran incompleteness tests using the central star
(NICMOS 145, AB=22.1) taken from 5 different individual exposures,
all taken at different dither positions. These have only 1/120 the
exposure time of the final image and therefore have lower signal-to-noise
ratios. We found no substantial differences in detection
rates, other than as the star is dimmed to AB$=$28, it is missed
most frequently in the upper right hand quadrant of the detector,
the least sensitive region of the image.
In the final combined image, we found no obvious location dependence.
We detect the star 95\% of the time as it is dimmed to AB$=$27, and
then the detection rate drops rapidly to 50\% complete at AB$=$28.5.

\section{Source Reliability Tests}
\label{rel}

Even though the listed catalog does not contain objects with signal to noise ratios less than 2.5 there can be false detection still in the catalog.   As indicated in Section \ref{cat-ref} the catalog indicates the degree of coincidence between the various subcatalogs that make up the total catalog.     This data is provided as an aid in discerning the reality of the sources.  Any statistical study of these results should utilize the test flag indices of Table \ref{cat} carefully along with the completeness and reliability results discussed here and in Section \ref{comp} and summarized in Table \ref{cr}.  See section \ref{crt} for a discussion of this table.  

\subsection{Half Data Reductions}
\label{half}
Our primary test of the reliability of observed sources utilizes two independent images formed from subsets of the integrations in each filter. The two images contain the even and odd numbered integrations from a sequential numbering of the integrations after removal of images with excess cosmic ray persistence. Since there are three images per orbit this technique insures that orbits are mixed between the groups and that each group has an equal mix of images observed at different times during the orbit. The width of the histograms of pixel values in the half images are a factor of $\sqrt{2}$ wider than the full image histograms.  This is a good indication that the width of the histogram in Fig. \ref{Thompson.fig5.ps} is due to noise rather than faint sources.

These half data reductions are the primary tests as they represent truly independent sets of data that measure the same quantity.  Although useful, the coincidence between the KFOCAS and SExtractor catalogs are not measurements of two independent data sets.  The coincidence between the objects detected in the various NICMOS and WFPC2 filter sets are again useful but they are not measuring the same quantity.  As with the completeness tests, the checks on the KFOCAS and Sextractor image catalogs are carried out independently.  With slight modifications, however, the logic of the tests is essentially identical.  
 
\subsubsection{Logic of the half catalog tests}
\label{logic}

Our goal is a measurement of the probability that a detected source with a given magnitude range is real. To facilitate the comparison between KFOCAS and SExtractor tests we both utilize aperture magnitudes in a $0.6\arcsec$ diameter aperture. We start by grouping sources into half magnitude bins centered on integer and half magnitudes.  Our analysis method then utilizes the statistics of objects detected in both, only one, or neither, of the independent half catalogs for each aperture magnitude bin.  We consider objects as being present in both catalogs if their centroids are within 0.25 arc seconds (2.5 drizzled pixels) of each other. 

From the completeness studies described in Section \ref{comp} we determined the probability $P_{A,B}(j,k)$ that an object in a  magnitude bin j is recovered in a bin k where A or B refers to one of the half catalogs.  The completeness is then  C$_{\rm A}(j) = {\displaystyle\sum_k} {\rm P_A}(j,k)$ 
and similarly for image B.  Let N$_{\rm R}(j)$ be the number of real objects whose true magnitudes lie within bin j. In addition, due to noise fluctuations (both Poisson and non-random), there will be some false objects detected in bin $j$.  Let f$_{\rm A}(j)$ be the probability that any object in 
image A in bin $j$ is a false detection and let N$_{\rm A}(j)$ 
be the number of objects found in bin $j$ on image A.
Then 
\begin{equation}
\label{sext_equ1}
{\rm N_A}(j) = \sum_k {\rm N_R}(k) \times {\rm P_A}(k,j) + {\rm f_A}(j) \times {\rm N_A}(j)
\end{equation}
and 
\begin{equation}
\label{sext_equ2}
{\rm N_B}(j) = \sum_k {\rm N_R}(k) \times {\rm P_B}(k,j) + {\rm f_B}(j) \times {\rm N_B}(j).
\end{equation}

We then count the number of objects $N_{AB}(i,j)$ which appear in common
in both half images A and B, (ie, which agree in position to within 0.25" of
each other) and which have  measured magnitudes in A which place them
in bin i, and  measured magnitudes in B which place them in bin j. Then

\begin{equation}
\label{sext_equ3} 
{\rm N_{AB}}(i,j) = \sum_k {\rm N_R}(k)\times {\rm P_A}(k,i) \times {\rm P_B}(k,j) \end{equation}

Strictly speaking, we should add to equation \ref{sext_equ3} a term which represents the
number of times that a false detection in both half images will be coincident to within
0.25" and land in the two magnitude bins in question. In practice this number is small compared to one, and we neglect it.

If  $N_{bins}$ is the number of magnitude bins, then  equations \ref{sext_equ1} and \ref{sext_equ2} hold for the $N_{bins}$ values of j, and in equation \ref{sext_equ3} for the $ N_{bins} \times N_{bins}$ combinations of i and j. The unknown quantities are
the $N_{bins}$ values for the number of real sources with true flux placing
them in bin $N_R(k)$ along with the $N_{bins}$ estimates for the probabilities
$f_A(j)$ and $f_B(j)$ that a given source in bin j is not a real source. Obviously, the system is over--determined. This is to be expected since equations \ref{sext_equ1}, \ref{sext_equ2} and \ref{sext_equ3} are just discrete representations of integral equations describing the
observed number count distribution from which we are trying to recover the
true distribution, taking into account losses, false sources and errors between
the true and measured magnitudes due to noise. 

A simplification of the preceding equations which is useful for illustrative
purposes and actual calculations in some cases comes from ignoring the cross
terms and letting the completeness in any bin be equal to C independent of
which half catalog is addressed.  This eliminates the cross terms in equations \ref{sext_equ1}, \ref{sext_equ2}, and \ref{sext_equ3}.

In that case, we obtain for each bin, 

\begin{equation}
\label{simp1}
N_{R} = N_{AB}/C^2
\end{equation}
and 
\begin{equation}
\label{simp2}
f_{A} = 1 - N_{AB}/(C \times N_{A})
\end{equation}
with a similar expression for $f_B$.

\subsubsection{Negative Image Tests}
\label{negim}
As described below, we are limited in the applicability of the full formalism described
above due to small number statistics in the observed number of objects, which
can lead to reliability estimates greater than one. An alternative procedure is to
multiply the final images by -1.0 and search for ``detections'' of objects in these
negative images. This assumes that the noise properties of the images are the same for
negative excursions as for positive ones. This is not in general true since, for example,
cosmic rays which are not completely removed have no counterpart in the negative image.
In the case of the NICMOS HDF images, however, not only are the cosmic rays removed
fairly effectively within each frame as a consequence of the non-destructive readout,
but the very large number of dithered frames making up our final images reduces any
residual cosmic rays by a further larger factor.
Unfortunately, we have found that this method does not appear to be well-suited for
the KFOCAS extractions for reasons associated with edge effects
near the  large negative ``holes'' in the counts
produced by the bright real sources in the negative images. However, this method does
seem to yield useful results for the SExtractor algorithm. We now discuss the particular tests actually carried out on the KFOCAS and SExtractor half catalogs.

\subsection{KFOCAS Half Catalog Tests}
\label{krel}

The KFOCAS analysis of each half image used the same parameters as the total image analysis.  Since the images are in units of photon flux the half images have the same signal strength for true sources but have a higher noise.  Unlike SExtractor, the $1\sigma$ noise level for KFOCAS is an input parameter.  Retention of whole catalog input parameters results in an input $1\sigma$ noise value that is a $\sqrt{2}$ lower than the noise in the half catalog.  The half image KFOCAS analysis then detects more sources since more random noise fluctuations appear above the detection threshold.  Bright true sources should be detected in both images.  Faint sources of course could be detected in only one or even neither of the half images.  Each source is marked in the catalog test column as to whether it appeared in both, only one or none of the half images. 

In practice for the KFOCAS source we utilize the simplified formalism described in equations \ref{simp1} and \ref{simp2} of Section \ref{logic}.  In particular we note that the reliability in either half catalog is $1 - f_{A}$ or $1 - f_{B}$ so we can say that the reliability r is

\begin{equation}
\label{simp3}
r_{A} = N_{AB}/(C \times N_{A})
\end{equation}

Since all of the quantities on the right side of equation \ref{simp3} are known r can be calculated using the values of C previously determined.  However, when this is formally carried out the values of r for some magnitude bins become greater than 1 due to a low value of C for that bin or small number statistics.  Since the true value of the completeness can never exceed one we can get a robust lower limit on r by
setting C equal to 1 and noting that 

\begin{equation}
\label{simp4}
r_{A} \geq N_{AB}/N_{A}
\end{equation}
with again a similar equation for $r_{B}$.  This equation depends solely on the
ratio of the number of sources detected in both catalogs to the number detected in one of the half catalogs.  The final reliability for a magnitude bin is just the average of $f_{A}$ and $f_{B}$.  This reliability is appropriate for the half catalogs.  The signal to noise in the whole catalog is a factor of $\sqrt{2}$ higher.  This corresponds the objects in the half catalog that are the same factor brighter.  This is an offset of 0.376 magnitudes, therefore, the calculated reliability numbers in the half catalog are appropriate for sources that are 0.376 magnitudes fainter in the whole catalog. 

Table \ref{cr} contains the results of these tests under the section marked KFOCAS.  The completeness values listed in the table are the values found from the analysis in Section \ref{kcomp}.  The reliability numbers are the numbers from the above calculation adjusted to the appropriate aperture magnitude for the total catalog. These values are in general lower than those calculated using the measured values for completeness in equation \ref{simp3}.

\subsection {Sextractor Half Catalog Tests}
\label{srel}

For both the  whole image and half images, we use the SExtractor parameters given
in table~\ref{t_sextparam} but determine the completeness
independently for  the half images using the same procedure described in  
\ref{comp}.
 
As noted above, the noise properties in the half images scale almost exactly
as expected, so that the false detection rate at a given magnitude bin in the
half images should be applied to a magnitude bin in the whole image fainter by
$1/\sqrt{2}$ lower in the flux, or 0.376 magnitudes.
We then use the mean of the two false detection
rates determined from the half images for the estimate of the
false detection rate at this slightly fainter magnitude.

As described in \ref{logic}, the system of equations 
\ref{sext_equ1}--\ref{sext_equ3} for ${\rm N_R}(k)$ is over-determined and we 
determined these values by a least squares fit to the observed values ${\rm N_{AB}}(i,j)$. Equations \ref{sext_equ1} and \ref{sext_equ2} then give the false detection rates for
the two half images, and use the mean of these determinations for the whole image
as explained above.

In practice, as already discussed in \ref{krel}, the small number of sources actually
detected in common in the two half images result in uncertainties in the reliability
estimates for magnitudes at which the completeness is near unity.
We have also estimated the reliability of the detections by the negative image method
described in \ref{negim}. The objects which SExtractor finds using this technique do
not seem to occur preferentially near the ``holes'' associated with the negative sources
but occur in the higher signal to noise regions, as expected, so that SExtractor does
not seem subject to the same degree to the edge effects described in \ref{negim}
for KFOCAS.

\subsection{The Completeness and Reliability Table}
\label{crt}
The completeness and reliability table, Table \ref{cr}, summarizes the results of our 
tests described above.  Columns 2--8 refer to the KFOCAS reductions only and the last three refer to the results from SExtractor.  As described in \ref{kscomp}, there is a systematic difference between aperture magnitudes
measured by KFOCAS and SExtractor which becomes significant for objects fainter 
than $\approx 28.0$, as shown in Figure \ref{Thompson.fig8.ps}. Thus, in the first column, the magnitude (Mag.) 
is the aperture AB magnitude measured by KFOCAS for the KFOCAS reductions and by
SExtractor for the SExtractor reductions. The width of the magnitude bin is 0.5 magnitudes centered on the value in the magnitude 
column. The signal to noise ratio (S/N) is the average signal to noise ratio for all 
objects in the magnitude bin.  The columns labeled
$Cs_{16}$, $Cl_{16}$, $Cs_{11}$ and  $Cl_{11}$ are the completeness numbers for the 
KFOCAS reduction listed in order for small and large objects in the F160W filter and 
the F110W filter, respectively.  Next are the reliability numbers for the F160W and 
F110W filters where no discrimination has been made between small and large objects.  
Completeness and reliability SExtractor results for the F160W filter comprise 
the last three
columns, where the column labeled $R_{co16}$ uses the full half image formalism, while
the column labeled $R_{neg16}$ uses the negative image technique. It should be
emphasized that the reliability estimates at the faint end of the table are
subject to considerable uncertainty. However the results all seem to indicate
a fairly low incidence of false detections $\sim 5-15 \%$ at magnitude $\sim27.5$ but
this incidence rises steeply at fainter magnitudes, while the completeness is of
order $\sim 80-90\%$ at magnitude 27.5, of order $\sim 70-75\%$ at magnitude 28.0 and
falls rapidly beyond that point.  KFOCAS appears to lose more objects due to merging with other objects than SExtractor particularly for bright objects. This is probably  the cause of the less than $100\%$ completeness at bright magnitudes for the KFOCAS reductions.  The low number of objects in the brighter bins limits the accuracy of 
the measurements and differences of $ \pm 5\%$ should not be considered significant.

Our discussion of the differences in source extractions in section \ref{kscomp} clearly indicates that the IDL and IRAF images are essentially identical and that the magnitude differences in fig. \ref{Thompson.fig8.ps} are solely due to differences between the two extraction programs, KFOCAS and SExtractor.  There are also differences in the number of detections between the two programs. Running SExtractor on the IDL image with the same set of parameters used for the SExtractor analysis of the IRAF images we find 284 galaxies, somewhat less than the 356 found on the IRAF image. On the other hand there are a total of 350
objects selected by KFOCAS from the whole IDL F160W image, also more than those
found by SExtractor on the IDL image. The total number of objects found in the F160W image by KFOCAS and in the IRAF image by SExtractor are nearly identical.

Inspection of table \ref{cr} indicates the range of reliability and completeness measures returned by the two methods.  In general the reliability and completeness measures from the KFOCAS analysis fall below those determined via the SExtractor analysis.  This is particularly true when you consider the difference in faint magnitudes discussed in section \ref{kscomp}. This indicates that at the faintest magnitudes the KFOCAS numbers should be compared with the SExtractor numbers for sources with SExtractor magnitudes nearly a magnitude brighter than the KFOCAS magnitude.  Part of this difference in reliability is due to the KFOCAS numbers being lower limits on the reliability as discussed in section \ref{krel}.  Another part of the difference, however, is due to the uncertainty inherent in these calculations and users of this catalog should be aware of them.  Our net philosophy is to be aggressive in identifying potential sources but to be relatively conservative in calculating their reliability and completeness.

\section{GALAXY COUNTS}

As with the original optical catalog of \cite{will96} it is not our intention in this paper to discuss scientific results.  A commonly used statistic, however,
is the differential number count of galaxies.  Fig. \ref{Thompson.fig10.ps} presents this statistic for the region of the HDF covered by our catalog.  The galaxy counts in number per magnitude per square degree have been divided into half magnitude bins.  If the object radius is less than $0.3 \arcsec$, the aperture magnitude is used.  If the object radius is greater than this the isophotal magnitude is used.  The number counts for the F814W filter in the same area as determined from the optical catalog are also shown for comparison.  The same division between aperture and isophotal magnitudes are used for the F814W data. The counts for the WFPC2 F814W objects includes all of the WFPC2 catalog objects in the NICMOS field, not just those in common with NICMOS objects. There are no aperture corrections applied to this data in order to facilitate comparison with the statistics presented in \cite{will96}. It should be noted that the area covered by the NICMOS image is very small. For a value of $H_{o} = 50$ and a value of $\Omega_{o} = 1$ the sides of the NICMOS image are on the order of 250 kpc for redshifts greater than 1.  This is about 10 times smaller that the typical diameter of a region forming a single galaxy from Cold Dark Matter simulations (\cite{stz98}). Drawing any cosmological conclusions from this small sample may be very suspect.  Following the discussion in \cite{will96} we have not eliminated split objects from the number counts.  Except at very bright magnitudes we do not expect this to significantly affect the statistics.

\placefigure{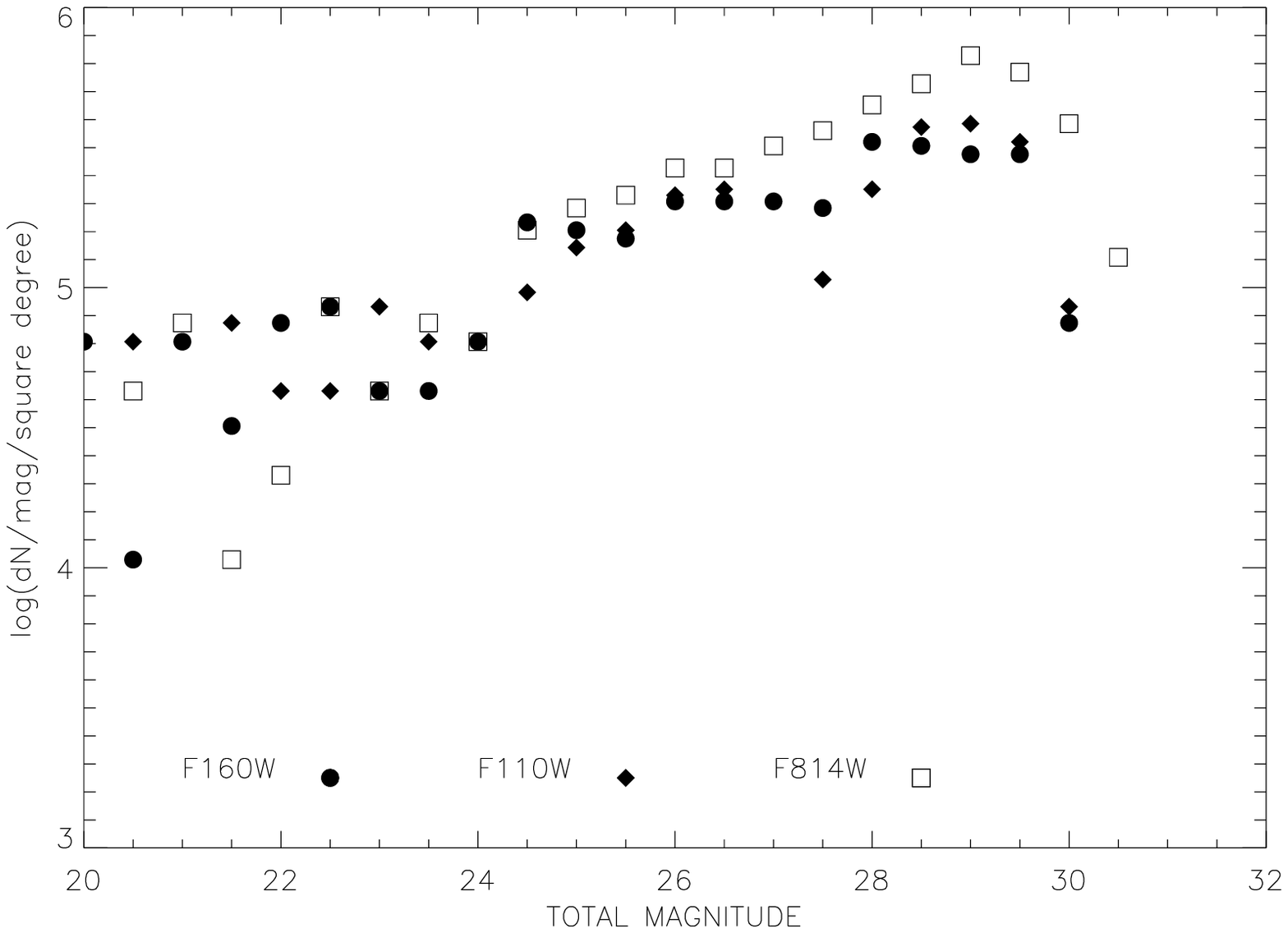} 

\section{CONCLUSIONS}

The NICMOS observations of the Hubble Deep Field add significant value to the existing data by providing improved wavelength coverage and access to objects that are either too heavily extincted or too highly redshifted to be visible in the original optical catalog.  This paper is designed to be a reference source for the use of this data in various areas of research.  Future papers will discuss various aspects of the significance of these data.

\section{ACKNOWLEDGEMENTS}
Many people contributed to the NICMOS observations of the Hubble Deep Field.  The entire NICMOS Instrument Definition Team contributed to the success of NICMOS and participated in the decision to allocate a large fraction of the team's guaranteed time to this effort.  We wish to thank Mark Dickinson for help with the KFOCAS reduction techniques and Andy Fruchter for his aid in implementing the Drizzle image procedure.  Andy Lubenow spent many hours refining our observation plan to handle the single guide star acquisition.  Chris Conner and the Lockheed MOSES group went to extraordinary efforts to keep the gyro biases updated to ensure good pointing under single guide star tracking.  Zolt Levay provided invaluable assistance in the preparation of the images for publication and Dr. Bertin provided the SExtractor software and quick response to inquiries.  The schedulers at STScI diligently worked to minimize the impact of SAA crossings. LS-L thanks Lin Yan and Patrick McCarthy and  RJW thanks David Koo for very useful discussions on incompleteness testing and galaxy surveys. This work was supported by NASA grant NAG 5-3043 and the observations were obtained with the NASA/ESA Hubble Space Telescope operated by the Space Telescope Science Institute managed by the Association of Universities for Research in Astronomy Inc. under NASA contract NAS5-26555.

\clearpage

\begin{figure}
\caption{The normalized total filter response functions for the NICMOS F160W and F110W filters. These response functions include all of the color dependent terms including the detector quantum efficiency.}
\label{Thompson.fig1.ps}
\end{figure}

\begin{figure}
\caption{f160w image}
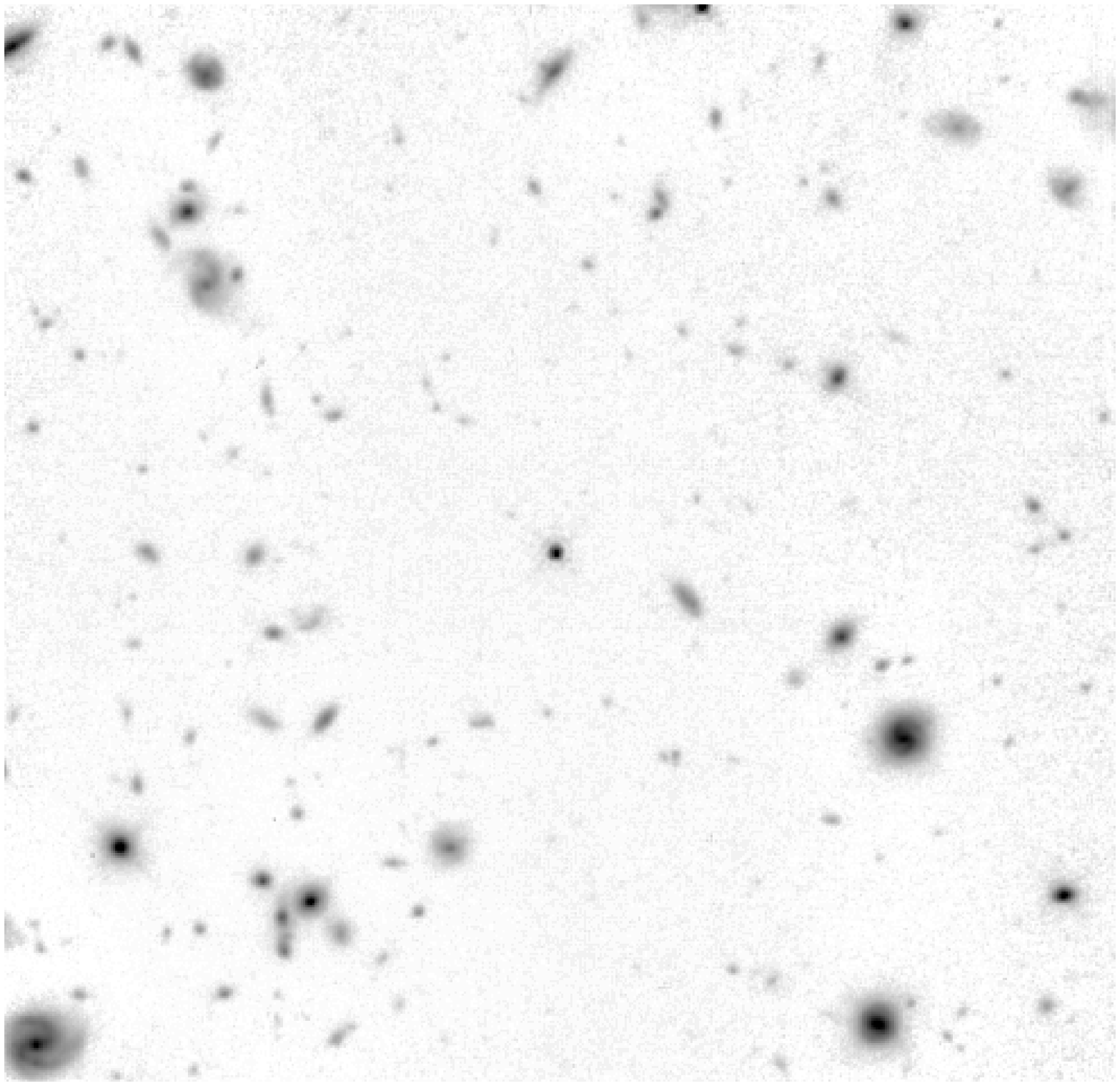
\label{Thompson.fig2.ps}
\end{figure}

\begin{figure}
\caption{f110w image}
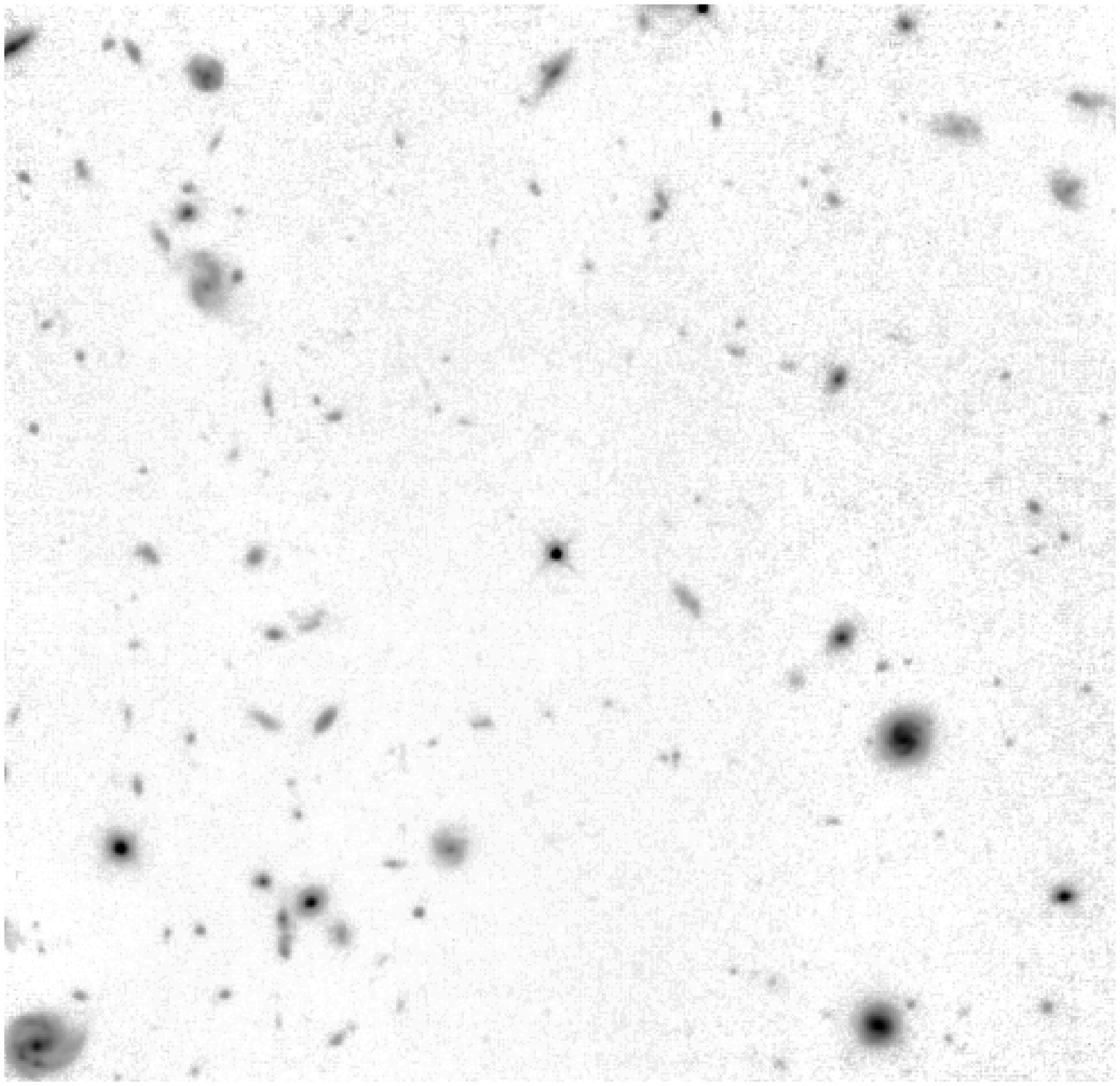
\label{Thompson.fig3.ps}
\end{figure}

\begin{figure}
\caption{composite color image}
\label{Thompson.fig4.ps}
\end{figure}

\begin{figure}
\caption{Histograms of the pixel values in the F160W and F110W mosaic frames are shown. Gaussian fits are overplotted as dotted lines.  The measured $1\sigma$ noise levels per pixel are $2.2 \times 10^{-4}$ ADUs per second ($6.1 \times 10^{-10}$ Jy, 31.9 AB mag) for the F160W image and $2.5 \times 10^{-4}$ ADUs per second ($7.7 \times 10^{-10}$ Jy, 31.7 AB mag) for the F110W image.  The true noise levels are about a factor of 2 higher than these values due the correlated noise produced in the drizzling process. }
\label{Thompson.fig5.ps}
\end{figure}

\begin{figure}
\caption{These are the F160W source detectability contours for the region included in the catalog.  The contours inside the F160W area cover a factor of 3.77.  The contour levels are $5\%$ of this range. The regions with the highest detectivity are in the center and lower left.}
\label{Thompson.fig6.ps}
\end{figure}

\begin{figure}
\caption{These are the F110W source detectability contours for the region included in the catalog. The contours inside the F110W area cover a factor 4.75.  The contour levels are $5\%$ of this range.  The regions with the highest detectivity are in the center and lower left.}
\label{Thompson.fig7.ps}
\end{figure}

\begin{figure}
\caption{A comparison of KFOCAS and SExtractor aperture magnitudes before the catalog cut at $2.4\sigma$.  The first panel shows the difference in SExtractor magnitudes determined from the IDL and IRAF images. The second and third panels indicate the differences between the two source extraction programs, KFOCAS and SExtractor. The last panel is a good guide in comparing the SExtractor determined magnitudes in the catalog with those determined by KFOCAS}
\label{Thompson.fig8.ps}
\end{figure}

\begin{figure}
\caption{The signal to noise ratios for faint sources relative to the KFOCAS aperture magnitudes for the F160W and F110W filters. This figure contains all of the extracted sources not just the catalog sources.}
\label{Thompson.fig9.ps}
\end{figure}

\begin{figure}
\caption{Differential galaxy counts as a function of AB aperture magnitude only of the sources appearing in the sigma limited catalog.}
\label{Thompson.fig10.ps}
\end{figure}
\begin{deluxetable}{cccccccc}
\tiny
\tablecaption{Basic Parameters of the NICMOS Hubble Deep Field \label{hdf}}
\tablewidth{0pt}
\tablehead{
\colhead{RA (J2000.0) } & \colhead{DEC (J2000.0)} & \colhead{Orient} & \colhead{Total Field} & \colhead{F110W} & \colhead{F160W} & \colhead{G096} & \colhead{G141}
} 
\startdata
$12^{\rm h} 36^{\rm m} 45.129^{\rm s}$ & $+62\arcdeg 12\arcmin 15.55\arcsec$ &$261.851\arcdeg$ &$57.87 \times  61.34\arcsec$ &49 orbits &49 orbits&12  orbits&15 orbits
\enddata
\end{deluxetable}

\begin{deluxetable}{crr}
\small
\tablecaption{KFOCAS Parameters \label{kfoc}}
\tablewidth{0pt}
\tablehead{\colhead{KFOCAS Parameter} & \colhead{F160W} & \colhead{F110W}}

\startdata
Magnitude zero point &35.3 &35.186\nl
Catalog magnitude limit &100. &100.\nl
Radius of fixed circular aperture &6.0 &6.0\nl
Sigma of sky &20 &22\nl
Sigma above sky for detection &2.5 &3.0\nl
Sigma below sky for detection &20.0 &20.0\nl
Minimum area for detection &2 &2\nl
Significance for evaluation and splitting &0.15 &0.15
\enddata
\end{deluxetable}

\begin{deluxetable}{ll}
\tablecaption{SExtractor Parameters \label{t_sextparam}}
\tablewidth{0pt}
\tablehead{
\colhead{Parameter} & \colhead{Value} } 
\startdata
DETECT\_MINAREA &  2              \\ 
DETECT\_THRESH  &  2.15           \\ 
ANALYSIS\_THRESH&  2.15 \\
FILTER\_NAME    &  gauss\_2.0\_3x3.conv \\
CLEAN          &  N               \\ 
MASK\_TYPE      &  CORRECT         \\ 
WEIGHT\_TYPE    &  MAP\_WEIGHT\\
PHOT\_APERTURES &  6  \\
GAIN           &  6.5       \\ 
PIXEL\_SCALE    &  0.1  \\
SEEING\_FWHM    &  0.2  \\
BACK\_SIZE      &  64      \\ 
BACK\_FILTERSIZE&  3       \\ 
BACKPHOTO\_TYPE &  LOCAL   \\ 
BACKPHOTO\_THICK&  24  \\
MAG\_ZEROPOINT  & 22.80 (F160W)\\
MAG\_ZEROPOINT  & 22.68 (F110W)\\
\enddata
\end{deluxetable}%

\end{document}